# Single Event Upsets characterization of 65 nm CMOS 6T and 8T SRAM cells for ground level environment

**Daniel Malagón, Gabriel Torrens, Jaume Segura, Sebastià A. Bota**

Grup de Sistemes Electrònics, Departament de Física. Universitat de les Illes Balears,

Palma de Mallorca, Spain.

*E-mail: gabriel.torrens@uib.es*

*Abstract.-* We present experimental results of the cross-section related to cosmic-ray irradiation at ground level for minimum-sized six-transistors (6T) and eight-transistors (8T) bit-cells SRAM memories implemented on a 65 nm CMOS standard technology. Results were obtained from accelerated irradiation tests performed in the mixed-field irradiation facility of the CERN High-energy Accelerator test facility (CHARM) at the European Organization for Nuclear Research in Geneva, Switzerland. A 1.45× higher SEU cross-section was observed for 6T-cell designs despite the larger area occupied by the 8T cells (1.5× for MCU). Moreover, the trend for events affecting multiple bits was higher in 6T-cells. The cross-section obtained values show that the memories have enough sensitivity to be used as a radiation monitors in high energy physics experiments.

## 1. Introduction

Radiation related effects have been a serious concern for circuits operating in harsh environments with presence of high-energy particles such as outer space, avionics, nuclear power plants or particle accelerators for high-energy physics research. However, as a consequence of the progressive size scaling of current solid-state devices, these effects also become a reliability concern in electronic systems performing critical applications even if operating at ground level environments [1].

Particle interactions with silicon atoms of an integrated circuit crystalline substrate are capable of generating electron–hole pairs whose drift and diffusion can induce disturbing transient currents when collected at specific circuit nodes. When the collecting node belongs to a combinational digital circuit, a Single Event Transient (SET) is generated, whereas if charge is collected at a memory cell internal node, it can modify its logic state leading to a corrupted data bit, causing a Single Event Upset (SEU) [2].

A SEU only happens when the collected charge exceeds a given threshold, usually known as critical charge, $Q_{crit}$. Such a $Q_{crit}$ value depends on process parameters, memory cell architecture, cell layout, supply voltage, logic state, and the transient characteristic of the perturbing current [3, 4]. SEUs can be further categorized as Single Bit Upset (SBU) when the state of a single cell or register is affected, or as a Multiple Cell Upset (MCU) when more than one cell or register are modified by the event. Note that a SET generated within a combinational block can be propagated through a number of logic gates and may eventually induce a soft error if captured by a register. The SEU itself does not permanently damage the affected device as it can operate normally after the disturbance [3].

With smaller device sizes and more aggressive design rules in nanometer technologies, the standard six transistors cell (6T), shown in Fig. 1, and widely used in SRAM memories has become more sensitive to device variations and more prone to functional failures than before. In fact, SRAM memories have been identified as the most susceptible devices to experience soft errors produced by radiation [5]. Alternative cell structures have been proposed to improve the performance of SRAM bit-cells. One of such alternatives adds a read port to 6T-cells, to get an 8T-cell (Fig. 1). 8T-cells achieve a clear improvement in read stability, being one of the weak aspects of 6T-cells, at the cost of an assumable area increase [6]. Cells with larger transistors have been proposed in [7], and cells with higher transistor numbers [8, 9] have also been proposed to reduce soft error occurrence at the expenses of a significant area increase or performance reduction, e.g. DICE, the SRAM cell used in space applications that consists of 12 transistors [10]. For this reason, conventional 6T-cells and, to a lesser extent, 8T-cells are still by far the most used.

The soft error cross section, σ, is a quantity that allows estimating a component sensitivity to a particle species. σ is related to the total count of soft errors $N_{SE}$ induced by irradiation with fluence Φ, by the following equation:

$$N_{SE} = \sigma\, \Phi \qquad (1)$$

It usually results more convenient to indicate the bit-cell cross section, $\sigma_{\text{bit-cell}}$, which can be obtained by dividing the total cross-section, σ, by the total number of memory bit-cells. Cross-section, typically used to quantify the bit-cell sensitivity to radiation, takes into account that only a fraction of all particles reaching the component will generate a soft error. Another used parameter is the soft error rate (SER), usually defined by the Failure-In-Time (FIT), one FIT is equivalent to 1 failure in $10^9$ device hours of operation, or by the Mean-Time-Between-Failure (MTBF) that is inversely related to FIT.

$$FIT = \frac{1}{MTBF} 10^9 \qquad (2)$$

The critical charge, computed from electrical simulations, is also used to quantify the robustness of a given memory cell, although it has been shown that there is not a monotonic dependence of $Q_{crit}$ with the measured SER of various bit-cell structures [11], and therefore neither with σ. This is because the response to radiation of a given circuit is also influenced by the layout geometry and charge collection efficiency of their sensitive nodes, factors that are not considered in electrical simulations used to determine $Q_{crit}$, as experimentally corroborated in [11].

In this context, the objective of this work is contributing to the characterization of an integrated SRAM memory on a 65 nm CMOS technology, designed with minimum sized 6T and 8T cells, by determining their cross sections at ground level. It is important to emphasize that few experimental results are available in the literature about the effects of radiation in 8T memories. Our experiments were performed using the CHARM facility at CERN. Additionally, since upset rate monitoring in SRAM memories was used to estimate the radiation fluence in High-Energy-Physics experimental areas [12], we include a discussion about the feasibility of using the tested memories as radiation monitors.

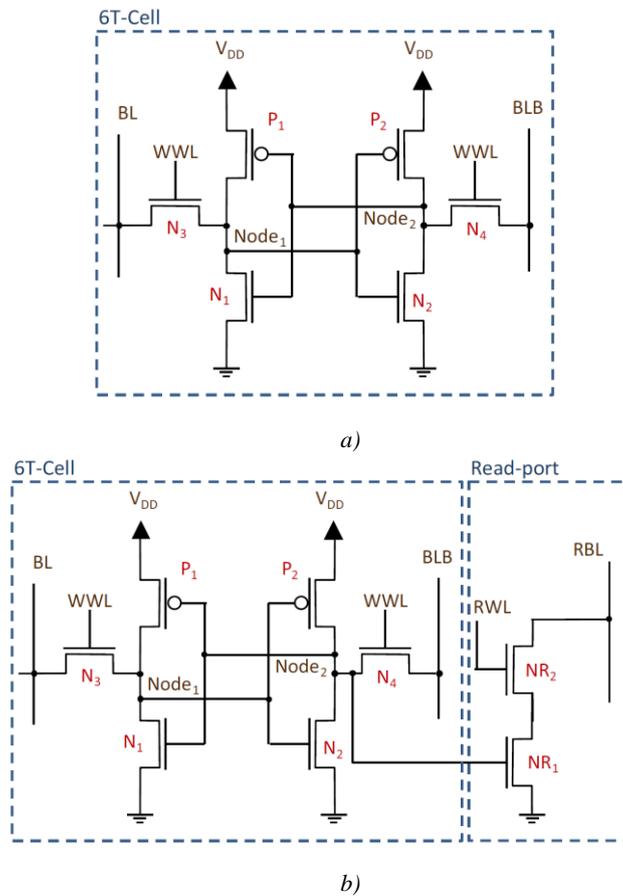

*Figure 1.* (a) Scheme of the standard 6T SRAM cell formed by pull-down transistors $N_1$ and $N_2$, pull-up transistors $P_1$ and $P_2$ and access transistors $N_3$, $N_4$. (b) 8T SRAM cell.

## 2. Mixed Field Environments

A high percentage of electronic components are subject to the influence of terrestrial cosmic rays, which are cosmic particles coming from outer space, and an additional component from solar ejections that reach the Earth. Most cosmic rays are trapped by the Earth's magnetic field and only a small fraction penetrate into the atmosphere where they produce cascades of secondary and tertiary particles (less than 1% of the primary flux reaches the sea level). At sea level, the environment consists of a mixture of neutrons, protons, pions, muons and other particle species [13]. Fig. 2 represents the corresponding ground level differential flux spectra that extends up to almost 10 GeV [14]; this flux varies with earth latitude and longitude increasing with altitude (at avionic altitudes the flux it is 200X higher than at sea level). It follows that the atmospheric radiation environment is characterized by its mixed nature, both in terms of particle types and energy interval, leading to the concept of mixed field environment.

Protons, neutrons, pions or kaons are hadrons, a class of subatomic particles made of two or more quarks, which can induce Single Event Effects because they produce inelastic nuclear interactions with matter. It is usually assumed that neutrons are the dominant cosmic ray by-products at ground level that impact electronics [13]. Neutrons can interact with the $^{28}$Si atoms producing secondary ions that may cause ionization tracks and create a sufficient number of electron-hole pairs to induce transient faults. These secondary particles can be generated anywhere in the semiconductor substrate and be emitted in any direction.

Besides neutrons, other particles such as protons, electrons, photons, positrons, muons and pions are present in the terrestrial cosmic rays. Electrons, positrons and photons do not usually induce soft errors in semiconductor devices, although recently, soft errors generated in nanoscale SRAMs irradiated with 20-MeV electrons have been observed as a consequence of a *nuclear reaction induced by the high-energy electron that lead to secondary particles* [15]. The reported cross-section was 3-4 orders of magnitude lower than those related to neutrons. The number of pions at ground level is negligible compared to neutrons, and so are its effects [13]. A charged muon loses its kinetic energy passing through silicon creating electron–hole pairs. When negative muons are quasi-stopped they must be captured by silicon nuclei releasing recoiling heavy nuclei with a simultaneous emission of light particles. The probability of generating soft errors increases when muon captures occur at the depth of the layer containing sensitive drains. Although this case corresponds to a reduced energy interval for the incoming muons [16], the main contribution to SER at sea level below 28 nm technologies comes from muons [17]. This is due to the critical charge decrease with scaling, and due to the higher muon flux compared with neutron flux at ground level.

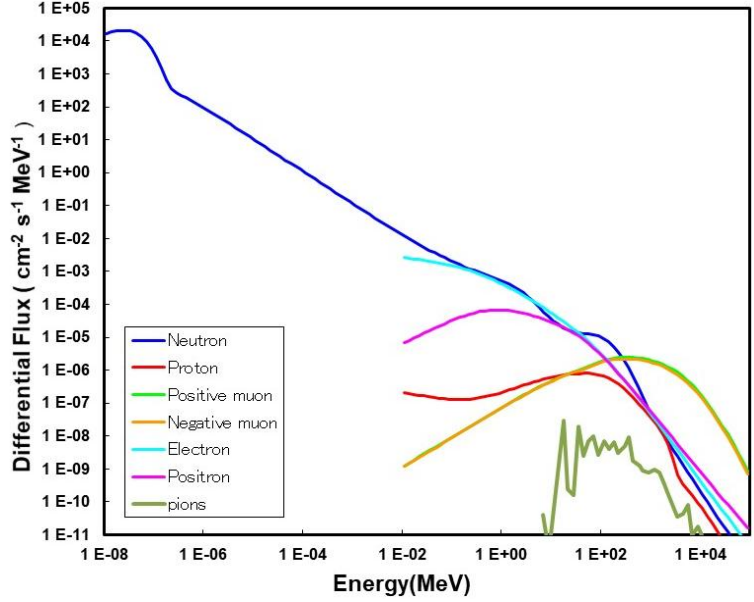

*Figure 2. Differential flux for atmospheric neutrons, protons, muons, positrons and electrons at sea level. Data computed using the PHITS-based Analytical Radiation Model in the Atmosphere [14] for Mallorca, Spain. Pions computed with QARM model [13].*

In High Energy Physics (HEP) environments like the Large Hadron Collider (LHC) at CERN, the accelerators themselves are radiation sources; hadrons are produced as by-products of intentional and non-intentional beam losses [18]. The spectra derived from Monte Carlo simulations for some LHC areas revealed the presence of hadrons, so they can also be regarded as examples of mixed field environments [18].

High Energy Hadrons (HEH) are defined as hadrons above 20 MeV. As a first approximation, it can be assumed that hadrons above 20 MeV are equally efficient in inducing SEEs due to their similar cross section in silicon [19]. The cross section as a function of energy for a given particle can be expressed as the product of a constant, $\sigma_{i0}$, and a weighting function $\omega_i(E)$, as in the next equation:

$$\sigma_i(E) = \sigma_{i0}\, \omega_i(E) \qquad (3)$$

For charged hadrons $\omega_i(E)$ is usually approximated by a step function with threshold at 20 MeV [20], while a Weibull function is used for neutrons.

$$\omega_n(E) = \left(1 - e^{-\left(\frac{E - E_{th}}{W}\right)^S}\right) \qquad (4)$$

Given the values of parameters $E_{th}$, S, W, it is usually obtained that $\omega_i(20\ MeV) \approx 1$.

In addition, the HEH fluence can be defined as:

$$\Phi_{HEH} = \int_0^T \int_{20MeV}^\infty \frac{d\varphi_{HEH}(E)}{dE} dE\, dt = \int_0^T \varphi_{HEH}\, dt \qquad (5)$$

Where the HEH flux, $\varphi_{HEH}$, is defined as the sum of the differential flux of hadrons above 20 MeV, and measured in particles·cm$^{-2}$·s$^{-1}$. However, it has been reported that below the 0.25 μm technology node, silicon bulk technologies exhibit a relatively high sensitivity to intermediate neutrons (neutron with energy range between 0.2 MeV and 20 MeV). It is explained by the contribution of secondary alpha particles coming from (n,α) reactions [21] together with the decrease of their critical charge due to the reduction of device sizes and biasing voltage. Given the presence of intermediate neutrons in both terrestrial and HEP environments, a correction must be introduced to account for their effect. Charged hadrons (protons, pions, kaons) below 20 MeV, either will not pass through the package or they will reach the sensitive volume with energy below the SEE threshold for indirect ionization [22].

An equivalent HEH flux, $\varphi_{HEHeq}$, *can be* defined as the sum of the HEH flux plus the intermediate neutron component, calculated by convoluting the energy differential neutron spectrum in the 0.2-20 MeV range, with $\omega_n(E)$ [21]:

$$\varphi_{HEHeq} = \varphi_{HEH} + \int_{0.2MeV}^{20MeV} \omega_n(E) \frac{dn(E)}{dE} dE \qquad (6)$$

*The HEHeq fluence is given by*

$$\Phi_{HEHeq} = \int_0^T \varphi_{HEHeq} dt = \Phi_{HEH} + \int_0^T \int_{0.2MeV}^{20MeV} \omega(E) \frac{dn(E)}{dE} dE\, dt \qquad (7)$$

It is typical to set a single cross section value for the HEH mixed-field contribution, $\sigma_{HEH}$, to relate the HEHeq fluence with the number of detected events $N_{SE}$:

$$N_{SE} = \sigma_{HEH} \Phi_{HEHeq} \qquad (8)$$

High-energy neutrons can lose energy by scattering with materials in the environment and ultimately reach the thermal equilibrium energy, becoming thermal neutrons [14]. At sea level the energy differential spectrum dn/dE presents a thermal peak at 25 MeV. It must be remarked that the iso-thermal and thermal part of the neutron spectrum is not capable of inducing soft errors originated by neutron-silicon interactions. However, they are capable of inducing soft errors in ICs because of the very large reaction cross section of these neutrons with $^{10}$B, an isotope commonly found in many IC processes as a substrate dopant and in the glass passivation layers. The interaction of thermal neutron with $^{10}$B produces a 1.87 MeV alpha and 0.84 MeV $^7$Li particles along with a 0.48 MeV photon [23]. Both α-particle and lithium nucleus release enough energy to produce sufficient electron-hole pairs to cause a cell upset. Although BPSG layers have been replaced in recent fabrication technologies, $^{10}$B remains still present as a dopant in the silicon bulk-substrate and the source/drain regions, in consequence, thermal neutron-induced single event upsets could still have an important contribution to SER in atmospheric applications [24].

## 3. Experimental details

### 3.1. SRAM test devices

The results presented in this work were obtained from standalone SRAM circuits fabricated with a 65 nm commercial CMOS technology, optimized for low power applications. The characterized SRAM consisted of an array of 256x64 6T bit-cells formed by minimum size transistors with $W_{N1,N2}=W_{P1,P2}=W_{N3,N4}=150$ nm (Fig. 1) for parasitic capacitance minimization. Despite of having a cell ratio $W_{N1,N2}/W_{N3,N4}=1$, cells were capable of being written, read, and could hold their state. A detailed analysis about the stability issues of these cells was reported in [25] and in [26]. The total area of the 6T SRAM core was 112.00 $\mu$m × 148.48 $\mu$m.

A second array of 256x64 8T cells was also implemented in the same silicon substrate by adding a read port of two nMOS transistors ($NR_1$ and $NR_2$), along with an extra read-word-line (RWL) and a dedicated read-bit-line (RBL) to the 6T cell (Fig. 1). The write-word-line signal (WWL) was used exclusively for write operations: in the 8T cell a read access did not disturb its contents, resulting in a cell with better noise margins [11]. The total area of the 8T SRAM core was 154,24 $\mu$m × 148,48 $\mu$m.

Registers in the memory control units that managed the access to the core bit-cells were implemented with triple voting modules mitigating the impact of hostile environments on their operation (especially when performing accelerated test experiments). The chip had three different power domains: a 3.3 V supply for the input-output (IO) cells, 1.2 V for the core logic (read/write logic such as decoders, sense amplifiers…), and 1.2 V supply for the SRAM core cells. Memory content could be accessed in parallel trough an 8-bit I/O module.

6T and 8T cells have two radiation sensitive nodes : $Node_1$ and $Node_2$, both belonging to the internal latch. Fig. 3 shows the physical layout of both bit-cells indicating the location of $Node_1$ and $Node_2$. Note that cell width is larger than its height. The layout area of 8T cells is 1.39 times larger than 6T cells area.

Fig. 4 illustrates the floorplan for a 4x4 block valid for both 6T and 8T arrays. This block was replicated 64 times along the vertical direction and 16 times along the horizontal direction to obtain the 256x64 memory. A logic upset could be caused by negative charge collection at the drain of the nMOS pull-down transistor connected to the cell node being at logic "1", or by positive charge collection at the drain of the pMOS pull-up transistor connected to the node being at logic "0". In both cases the SEU is originated by extra charge collection at a transistor in the OFF state. Usually the collection efficiency for holes in CMOS devices is smaller than for electrons. Moreover, the magnitude of $Q_{crit}$ required to flip the node being at "0" is higher than the corresponding $Q_{crit}$ required to flip the node being at "1", therefore the most sensitive regions for events induction are those closest to the pull-down transistor drain contact. The sensitive regions are circled in Fig. 4, assuming that all cells are initialized to logic-0; MCUs are expected to be more common when the sensitive regions are closely placed [27].

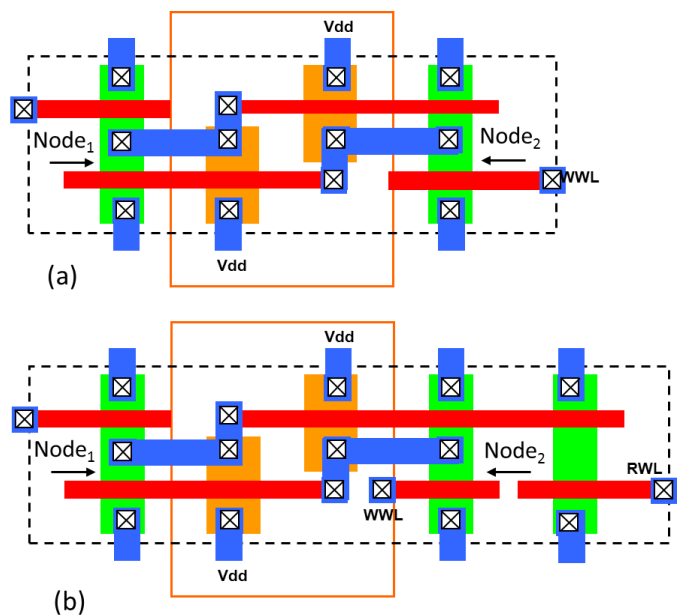

*Figure 3. (a) 6T-cell layout. (b) 8T-cell layout.*

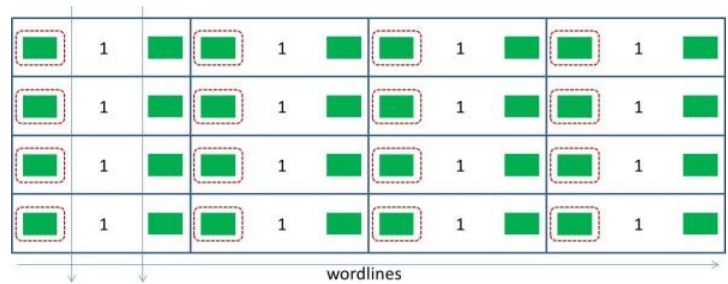

*Figure 4. Detail of a 4x4 block of the memory core, individual bit-cells are placed sharing their biasing contacts. The green rectangular shapes indicate pull-down transistor sensitive areas. The memory core is obtaining by repeating this block.*

### 3.2. HEH production at CHARM

Accelerated irradiation experiments were used to characterize the response of electronic component to a given radiation environment [13]. They consisted in the exposition of the device under test to an intense radiation source having a particle flux significantly larger than the flux under nominal conditions. This technique permits to significantly shorten the time needed to measure the soft error rate (or the cross section) that otherwise should be obtained from experiments that could last for several months. Data obtained from accelerated testing is then extrapolated to environment conditions through flux normalization. Table I reports the approximate annual HEH flux values in a variety of specific environments;

Table I. HEH fluxes for different radiation environments [28].

| Environment | $HEH\ flux\ (/cm^2/yr)$ |
|---|---|
| Sea level | $1\text{-}2 \cdot 10^5$ |
| Avionics | $\sim 2 \cdot 10^7$ |
| International Space Station Orbit | $\sim 7 \cdot 10^8$ |
| Polar LEO orbit (800 km) | $\sim 3 \cdot 10^9$ |
| LHC | $\sim 10^6\text{-}10^{12}$ |

In this context, the CHARM at CERN is a dedicated facility for electronic components accelerated testing, which can replicate a wide number of representative radiation environments such as sea level, space or HEP complexes [18]. The radiation environment inside the CHARM area is of mixed nature since the radiation within the test area comes from various particles type such as neutrons, protons, pions, etc, of a wide range of energies.

The CHARM mixed field is generated when a 24 GeV proton beam impinges a cylindrical target of cooper or aluminum (the beam can also be used without target). Depending on the target selection, the radiation field can be varied inside the test area, the device under test (DUT) can be placed in more than 14 standard test locations, each homogeneous at least in an area of one square meter, which can be selected based on the user requirements. Additionally, there are movable concrete and iron shielding elements that can be used to modulate the radiation field. Various mixtures of hadrons (protons, neutrons, pions and kaons), leptons (electrons, positrons and muons) and photons (x-ray and gamma) can be obtained in the CHARM test room. They range in energy from 24 GeV neutrons and protons (in places very close to the primary beam), to thermal neutrons [18].

The radiation spectra seen at each test position is determined from dedicated FLUKA Monte Carlo calculations [29]. A device placed in specific target locations is exposed to a mixed-radiation field capable of depositing a sufficient amount of energy in the silicon substrate to cause SEEs through indirect ionization mechanisms. However, it should be noted that it is not possible to separate the effects produced on the electronic component by different types of particle (each one with a different energy spectrum) present in the mixed field.

## 4. Results and Discussion

An accelerated HEH test was run in CHARM to characterize the reported memories in a mixed field representative of the atmospheric environment at sea level. The DUT was located in a rack place characterized by the spectrum shown in Fig. 5 to obtain a combination of particles with a spectrum as close as possible to the atmospheric environment.

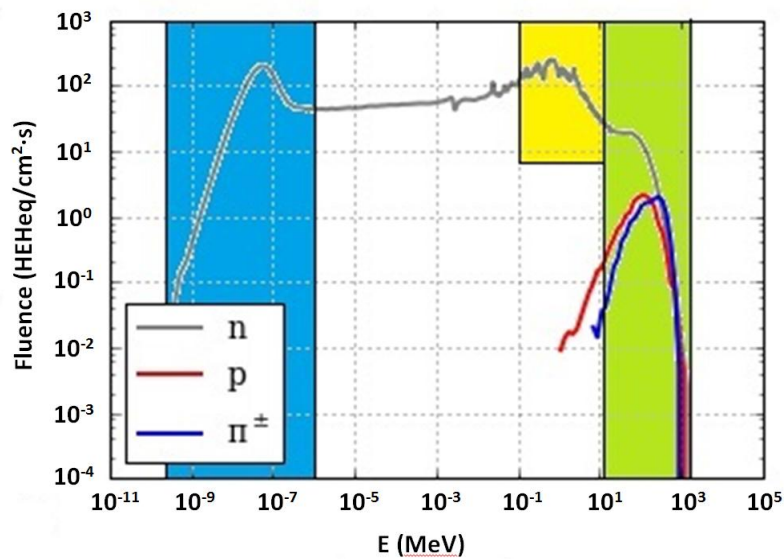

*Figure 5. FLUKA simulated lethargy spectra for the SEE-relevant hadrons at CHARM test facility [18] in the DUT location. The different shaded regions represent approximately thermal neutrons (blue), intermediate neutrons (yellow) and HEH (green).*

A specific test board, compatible with the CHARM set-up, was designed to properly bias the memories while setting up the adequate configuration to write and read the 6T and 8T memory cores during irradiation. SRAM control was achieved using Ethernet cables through an Altera FPGA development board controlled by a host PC through an USB port. The memories were initialized to all-0 (logic 0 was written in all memory cells) and their content was periodically read during irradiation. For further data analysis, the memory content was stored in the host PC each time that at least one bit change was detected, although due to an error in the interface link one bit of the output module eight bits was inaccessible limiting the test to 56x256 bit-cells. All experiments were performed with memory cores biased at 1.2 V. Inside the radiation room, the DUT was located in air as shown in Fig. 6.

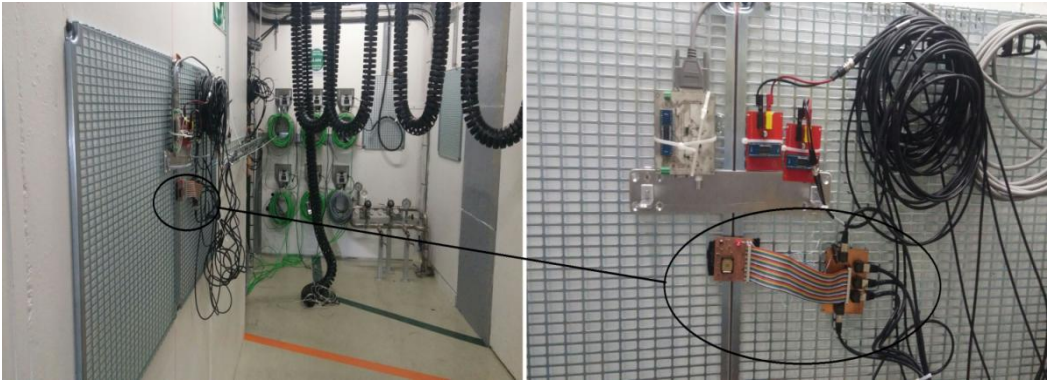

**Figure 6. DUT and Ethernet cables in CHARM radiation room.**

The estimated $HEH_{eq}$ flux at the test place was provided by the CERN Engineering Department and the Radiation to Electronics Group, by monitoring the proton beam intensity before interacting with a copper target and multiplying it by a proper conversion factor obtained from FLUKA simulations [29]. To illustrate this, Fig. 7 shows a monitoring of the estimated $HEH_{eq}$ fluence in the test place calculated with a period of 15 minutes. As shown, the beam flux was not constant and could even be momentarily interrupted during the experiment.

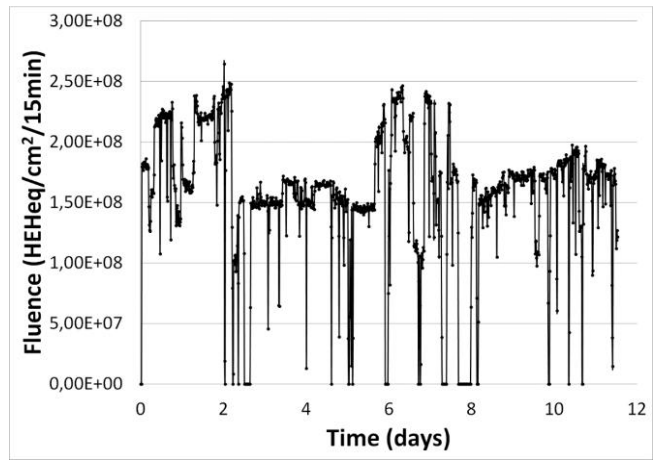

**Figure 7.** *$HEH_{eq}$ fluence estimated every 15 min. inside the CHARM room (sea level environment).*

After some preliminary tests performed under beam irradiation for adjustment, the sample was subjected to two irradiation periods. The first one had an overall duration of 60 hours (with an effective irradiation time of 27 hours, since due to maintenance tasks the flow was interrupted for about 33 hours while the sample was in test mode), and a second period of 12 hours. A total of 15 days elapsed between the irradiation periods, during which the memory remained inactive.

The overall fluence at the end of the first irradiation test was ~$2·10^{10}$ HEH$_{eq}$/cm$^2$ (It would take $1.3 \times 10^5$ years to reach this same fluence at sea level environment). A total of 214 events were detected in the 14K-bit 6T memory core and 147 in the 8T one. Particles interacting with the chip follow a Poisson process [30], consequently the relative uncertainty of the measurements is given by $1/\sqrt{N_{TOT}}$ being N the total number of SEUs [31]. This leads to a statistical counting error of about 7-8%.

From the irradiation data it was obtained that $\sigma_{SEU}$ = 7.5·10$^{-13}$ cm$^2$ for 6T-cells, and $\sigma_{SEU}$ = 5.2·10$^{-13}$ cm$^2$ for 8T-cells. We observed that the cross section of 6T cells was 1.45 times larger than that of 8T cells, although their sensitive nodes were similar having both practically the same critical charge (1.96 fC for 6T and 2.07 fC for 8T). This result reinforces once more that the critical charge is not capable of accurately characterizing the hardness of a cell. The reason for such a discrepancy could be associated with the fact that the nMOS transistors of the output port are in cut-off except during cell reading, and therefore during most of the experiment time they would act as a sort of guard drains by collecting part of the charge induced by the HEH interactions [32]. Note that, with the measured accelerated SER, it holds that a SER of 2.5 errors per year can be expected in a 16 Mbit 6T memory at sea level.

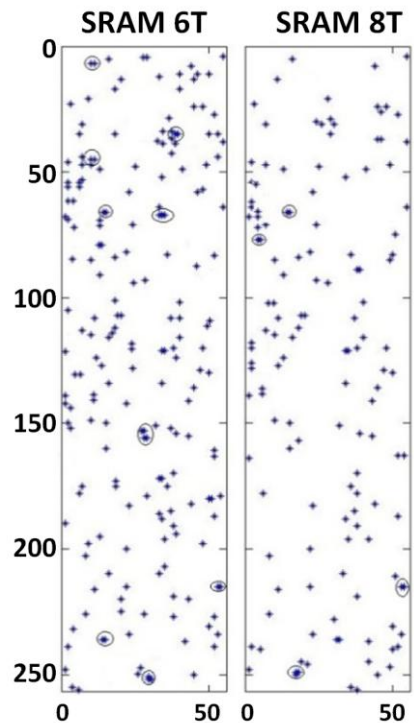

Figure 8. Bitmap of detected SEUs produced by HEH$_{eq}$ irradiation. Circles denote the presence of MCUs.

Table II. Multiple cell upsets.

|  | SEUs | MCU (2) | MCU (3) | MCU (≥4) | Total Events | Affected bits |
|---|---|---|---|---|---|---|
| SRAM$_{6T}$ | 205 | 9 | 0 | 0 | 214 | 223 |
| SRAM$_{8T}$ | 142 | 5 | 0 | 0 | 147 | 152 |

In nanometer SRAM chips, MCUs are likely to happen since SRAM cells are much smaller. Therefore, the number of affected bits in each event was also analyzed to identify the sensitivity to MCUs for the described cells. Figure 8 shows the placement of all the events measured highlighting the presence of MCUs (circles). 4,3% of the total events in 6T and 3,5% in 8T correspond to MCU (Table II). Detailed location of the affected bits in the memory confirmed that MCUs occurred in adjacent memory locations, pointing out that they were caused by the same recoil track.

Burst events [33], which can be explained by Single Event Functional Interrupt (SEFI) of the memory control circuitry were not detected during the experiment (probably due to the adoption of triple voting protection at the control unit).

A second 12 hours irradiation accelerated test was performed 15 days after the first irradiation sequence to validate the results of the previous experiment. The mean flux during the experiment was $2.0·10^5$ HEH$_{eq}$/cm$^2$·s corresponding to a total fluence of $8.6·10^9$ HEH$_{eq}$/cm$^2$. A total of 105 errors were detected in 6T cells corresponding to a cross section of $8.5·10^{-13}$ cm$^2$, and 105 errors in 8T cells corresponding to a cross section of $5.6·10^{-13}$ cm$^2$ (again 6T cross-section was 1.5X higher than 8T). These values, slightly higher than the previous ones, were in accordance to the experimental error. As in the previous test, we observed a 7.6% of events involving multiple bits in 6T and only 1.4% in 8T (Table III).

Table III. Multiple cell upsets.

|  | SEUs | MCU (2) | MCU (3) | MCU (≥4) | Total Events | Affected bits |
|---|---|---|---|---|---|---|
| SRAM$_{6T}$ | 97 | 3 | 3 | 2 | 105 | 120 |
| SRAM$_{8T}$ | 69 | 1 | 0 | 0 | 70 | 71 |

## 5. Summary and conclusions

A study to determine the effects produced by cosmic ray radiation at ground level in 6T and 8T memory cells fabricated on a 65 nm commercial CMOS IC technology were performed using an accelerated test in a hadronic mixed-field at CHARM facility at CERN.

The resulting cross sections are presented in Table IV along with the results obtained by other authors in related experiments. The cross-section values obtained in this work are in agreement with previously published values.

Table IV. *Reported cross-sections $cm^2 \cdot bit$*

|  | *Feature size* | *Particle energy* | *Cross-section* |
|---|---|---|---|
| S. Danzeca et al. [33] | 90 nm | Neutron-5 MeV-14MeV | $0.93 \cdot 10^{-14} - 1.09 \cdot 10^{-13}$ |
| D. Malagon et al. [34] | 65nm | Neutron-5.8-8.5MeV[6T-8T] | $3.86 \cdot 10^{-14}$ (6T)<br>$2.57 \cdot 10^{-14}$ (8T) |
| Wang et al. [35] | *130 nm* | *Proton-50-100 MeV* | *$1.02 \; 10^{-14}$* |
| Roche et al. [36] | *65 nm* | *Proton-10-60 MeV* | *$\approx 0\text{-}60 \; 10^{-14}$* |
| Heidel et al. [37] | *65 nm (SOI)* | *Proton-20-60 MeV* | *$0.8 \; 10^{-14}$* |
| Reed et al. [38] | *n.a.* | *Proton-24 MeV* | *$1.7 \; 10^{-14}$* |
| D. Lambert et al. [39] | 65 nm | Neutron-1-20 MeV | $2 \cdot 10^{-16} - 3 \cdot 10^{-14}$ |
| This work | 65 nm | Ground level mixed field spectrum | $7.6 \cdot 10^{-13}$ (6T)<br>$5.2 \cdot 10^{-13}$ (8T) |

Results in this work are also in line with [11] and [34], where it was also found a higher cross-section for standard 6T-cells when compared to 8T-cells under alpha particle irradiation (1.13X) and intermediate energy neutrons (1.5X). T*he statistical error related with the low number of detected events cannot guarantee that 8T cells are also more robust to MCU* although in general it is observed that an additional advantage of 8T over standard 6T is that they provide a better performance against radiation, probably caused by the presence of some isolation mechanism related to the presence of the read port in 8T-cells.

Electronic systems containing COTS components can already fail at hadron fluencies as low as $10^7$ cm$^{-2}$. These systems are usually installed in adjacent service areas of particle accelerators and other HEP installations, and thus exposed to a radiation field composed of various particles and energies. Radiation levels monitoring allows anticipating possible device degradation and identifying instantaneous failures of electronic equipment as caused by radiation. Therefore, monitoring systems capable of measuring radiation levels induced by HEH fluences ranging from $10^6$ to $10^{13}$ cm$^{-2}$ are required. Radiation monitors based on commercial SRAM components like Toshiba TC554001AF-70L with 0.4 um feature size and a cross section of 3.0 10$^{-14}$ cm$^{-2}$/bit for HEH, and the Cypress SRAM CY62157EV30 with a cross section of 3.8 10$^{-14}$ cm$^{-2}$/bit have been used in the the LHC RadMon radiation monitoring system at CERN. The results presented in this work also show that the tested 6T SRAM memories are excellent candidates to be used as HEH monitors. This is based, on the one hand, because we found in [34] that they were not sensitive to thermal neutrons (no events were observed after irradiating the samples with a fluence of 2·10$^{11}$ cm$^{-2}$) and, on the other hand, because their fluence resolution is better than the resolution of the commercial devices used in RadMon [12, 33]. Assuming a radiation monitor with a total memory of 16 Mbit (RadMon is based on 32 Mbit size), a fluence resolution of 8·10$^4$ cm$^{-2}$/count could be achieved providing good statistics measurements (>100 counts) of a fluence of ∼8·10$^6$ cm$^{-2}$.

**Acknowledgments**

The authors would like to thank R. García Alia and M. Brugger for their contribution to this work and the RME staff at CERN for his efficient technical assistance. Special thanks are finally due to S. Danzeca for his logistical support.

Caches," in *IEEE Journal of Solid-State Circuits*, vol. 43, no. 4, pp. 956-963, April 2008. https://doi.org/10.1109/JSSC.2007.917509.

[7] G. Torrens, I. d. Paúl, B. Alorda, S. Bota and J. Segura, "SRAM Alpha-SER Estimation From Word-Line Voltage Margin Measurements: Design Architecture and Experimental Results", in IEEE Transactions on Nuclear Science, vol. 61, no. 4, pp. 1849-1855, Aug. 2014. https://doi.org/10.1109/TNS.2014.2311697

[8] S. M. Jahinuzzaman, D. J. Rennie and M. Sachdev, "A Soft Error Tolerant 10T SRAM Bit-Cell With Differential Read Capability" in IEEE Transactions on Nuclear Science, vol. 56, no. 6, pp. 3768-3773, Dec. 2009. https://doi.org/10.1109/TNS.2009.2032090

[9] A. Islam, M. Hasan, "Variability aware low leakage reliable SRAM cell design technique", Microelectronics Reliability, vol. 52, no. 6, pp. 1247-1252, jun. 2012. https://doi.org/10.1016/j.microrel.2012.01.003

[10] T. Calin, M. Nicolaidis and R. Velazco, "Upset hardened memory design for submicron CMOS technology," in *IEEE Transactions on Nuclear Science*, vol. 43, no. 6, pp. 2874-2878, Dec. 1996. https://doi.org/10.1109/23.556880

[11] S.A. Bota, G. Torrens, J. Verd, J. Segura, "Detailed 8-transistor SRAM cell analysis for improved alpha particle radiation hardening in nanometer technologies", Solid-State Electronics, vol. 111, pp. 104-110, Sept. 2015. https://doi.org/10.1016/j.sse.2015.05.036

[12] G. Spiezia, P. Peronnard, A. Masi, M. Brugger, M. Brucoli, S. Danzeca, R, Garcia Alia, R. Losito, J, Mekki, P. Oser, R. Gaillard, L. Dusseau, "A New RadMon Version for the LHC and its Injection Lines," in *IEEE Transactions on Nuclear Science*, vol. 61, no. 6, pp. 3424-3431, Dec. 2014. https://doi.org/10.1109/TNS.2014.2365046

[13] J.L. Autran, D. Munteanu, Soft-errors: from particles to circuits, Taylor & Francis/CRC Press, Boca Raton (FL), (2015).

[14] T. Sato, "Analytical Model for Estimating Terrestrial Cosmic Ray Fluxes Nearly Anytime and Anywhere in the World: Extension of PARMA/EXPACS", *PLoS One* 10(12) (2015). https://doi.org 10.1371/journal.pone.0144679

[15] M. J. Gadlage, A. H. Roach, A. R. Duncan, M. W. Savage and M. J. Kay, "Electron-Induced Single-Event Upsets in 45-nm and 28-nm Bulk CMOS SRAM-Based FPGAs Operating at Nominal Voltage," in *IEEE Transactions on Nuclear Science*, vol. 62, no. 6, pp. 2717-2724, Dec. 2015. https://doi.org/10.1109/TNS.2015.2491220.

[16] J.M. Trippe, R.A. Reed, B.D. Sierawski, R.A: Weller, R.A. Austin, L.W. Messengill, B.L. Bhuva, K.M. Warren, B. Naramsimham, "Predicting the vulnerability of memories to muon-induced SEUs with low-energy proton tests informed by Monte Carlo simulations," *2016 IEEE International Reliability Physics Symposium (IRPS)*, Pasadena, CA, 2016, pp. SE-6-1-SE-6-6. https://doi.org/10.1109/IRPS.2016.7574642

[17] P. Li Cavolia, G. Huberta and J. Busto, "Study of atmospheric muon interactions in Si nanoscale devices" in Journal of Instrumentation, Volume 12, December 2017. https://doi.org/10.1088/1748-0221/12/12/P12021

[18] A. Infantino, R. G. Alía and M. Brugger, "Monte Carlo Evaluation of Single Event Effects in a Deep-Submicron Bulk Technology: Comparison Between Atmospheric and Accelerator Environment," in *IEEE Transactions on Nuclear Science*, vol. 64, no. 1, pp. 596-604, Jan. 2017. https://doi.org/10.1109/TNS.2016.2621238.

[19] R. G. Alia, M. Brugger, S. Danzeca, V. Ferlet-Cavrois, C. Poivey, K. Roed, F. Saigne, G. Spiezia, S. Uznanski, and F. Wrobel, "SEE Measurements and Simulations Using Mono-Energetic GeV-Energy Hadron Beams," in *IEEE Transactions on Nuclear Science*, vol. 60, no. 6, pp. 4142-4149, Dec. 2013. https://doi.org/10.1109/TNS.2013.2279690
16